\documentclass[amsmath,amssymb]{revtex4}

\usepackage[latin1]{inputenc}
\usepackage{graphicx}
\usepackage{dcolumn}
\usepackage{bm}
\usepackage{color}
\newcommand{\ket}[1]{|#1\rangle}

\begin{document}

\title{Delayed commutation in quantum computer networks.}
\author{Juan Carlos Garc\'ia-Escart\'in}
\email{juagar@tel.uva.es}
\author{Pedro Chamorro-Posada}
\affiliation{Departamento de Teor\'ia de la Se\~{n}al y Comunicaciones e Ingenier\'ia Telem\'atica. Universidad de Valladolid. Spain.}
\date{\today}
\begin{abstract}
In the same way that classical computer networks connect and enhance the capabilities of classical computers, quantum networks can combine the advantages of quantum information and communications. We propose a non-classical network element, a delayed commutation switch, that can solve the problem of switching time in packet switching networks. With the help of some local ancillary qubits and superdense codes we can route the information after part of it has left the network node. 
\end{abstract}
\maketitle

\section{Introduction}
Quantum information offers new possibilities in the fields of communication and computation \cite{NC00}. There are quantum algorithms capable of performing certain tasks with an efficiency that no known classical algorithm can match. Examples are Shor's algorithm for polynomial time factorization \cite{Sho97}, Grover's algorithm for unstructured database search \cite{Gro97}, and the Deutsch-Jozsa algorithm, that gives information on global properties of a function without an explicit evaluation of all the possible values \cite{Deu85,CEM97}. Quantum information theory extends classical information theory \cite{BS98} and includes analogues to classical Huffman codes \cite{BFG00}, and error correction codes \cite{Sho95,CS96,Got96}. Applications to quantum communication are especially interesting. Quantum cryptography proposals \cite{Tow97} give schemes for secure data transmission, and superdense coding \cite{BW92} allows to send more than one classical bit per transmitted qubit. 

The development of optical communication networks and the advanced state of the current photonic technology have permitted the experimental realization of cryptography \cite{MT95}, transmission of information in superdense codes \cite{MWK96}, and the teleportation of qubits \cite{BPW00}.

On the other hand, classical computer networks have witnessed an impressive progress in the last decades. The advent of multimedia data applications and the increased use of the Internet, telephone and other network services have brought an increasing bandwidth demand. The need for faster transmission rates has led to the progressive implantation of optical networks. Optical fibre technology is now used in the core of practically all long distance networks. 

The two basic models in network data transmission are packet and circuit switching \cite{Sta00}. In circuit switching, a path between source and destination is established at the beginning of the communication, and all the information uses the same series of network resources, that are solely dedicated to that connection. This is the model traditionally used in the telephonic network. In packet switching, the information is divided into groups of bits, packets, that are put into the network with a header that contains the destination. The route to the destination is not fixed, and all the elements in the path are dynamically assigned. The packets can even arrive out of order. That is the basic model of the TCP/IP protocol used for the Internet. In the last decades there has been a progressive shift towards packet switching, even for traditional circuit services (such as telephony, multimedia traffic, cellular systems \ldots). 

In packet switching it is particularly important how the decision on the path of the information is taken. The network nodes responsible for this task are called switches or routers. Depending on the contents of the header, the switch must decide the next node on the route. If the switch is not the final destination of the packet, some decision algorithm must tell to which of the nodes connected to the switch goes the packet. All these operations will introduce a \emph{processing time} that delays the packet and makes necessary, in most of the cases, some kind of buffer to keep the information of the packet. 

This a major drag for fast optical networks, where the processing is usually electronic, and requires an optical-to-electronic conversion. For this reason there has been a great interest in all-optical networks \cite{TZ99}. There are many proposals to reduce the constraints the routing imposes (see \cite{QY99,HA00}). 

The concepts of classical networking can be extended to quantum computers, and general switching methods using quantum gates have been designed \cite{TK02}. In this paper we propose a quantum switch that can reduce to zero the switching time by using a variation of superdense coding and some ancillary qubits in the switch. Applying delayed commutation techniques we no longer need long-term buffers, or delay lines able to entertain the packet and there is no processing time to contribute to the total packet delay, allowing for higher bit rates. 

Section \ref{notation} will define the notation and describe the quantum gates that appear in the following sections. Section \ref{concept} will introduce the main ideas behind delayed commutation, and what can and cannot be achieved using it in switching. Section \ref{simple} gives a quantum circuit that implements delayed commutation in a switch connecting two sources to two destinations. Section \ref{general} generalizes the result for any number of nodes. Finally, section \ref{discuss} compares the results to those of classical networks and explores future applications.   

\section{Notation and basic quantum gates.}
\label{notation}
Qubits are the basic information units in quantum information. They are the generalization of classical bits. As opposed to classical bits, that can be either 0 or 1, qubits can not only present the values $\ket{0}$ and $\ket{1}$, but also be in a superposition of states of the form $\ket{\psi}=\alpha\ket{0}+\beta\ket{1}$. Here $\alpha$ and $\beta$ are complex numbers and $|\alpha|^2$ and $|\beta|^2$ are the probabilities of measuring  $\ket{0}$ and $\ket{1}$ respectively. As they are probabilities $|\alpha|^2+|\beta|^2=1$.

\begin{figure}[h!]
\centering
\includegraphics{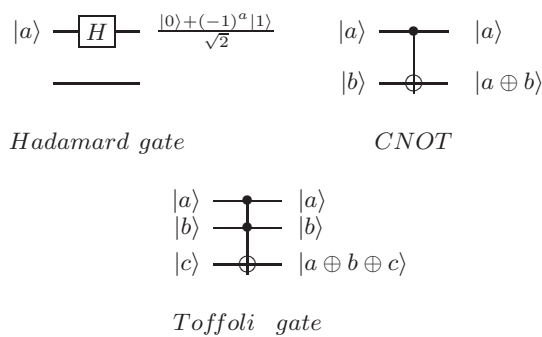}
\caption{Quantum gates used in the switch.} 
\label{gates}
\end{figure}

We can operate on qubits using quantum gates. In our switch we will use three kinds of gate, the Hadamard gate, the CNOT gate, and the Toffoli gate. The symbols for these gates can be seen in figure \ref{gates}.

A Hadamard gate takes the state $\ket{a}$ to $\frac{\ket{0}+(-1)^{a}|1\rangle}{\sqrt{2}}$. The CNOT gate leaves the state of the first qubit (control qubit) unchanged, and has at the output of the second qubit (the target qubit) the logical XOR of the contents of the input control and target qubits. If the value of the control qubit is $\ket{0}$ there is no change in the state. If it is $\ket{1}$ the value of the target qubit is flipped. The Toffoli gate is similar to the CNOT gate, but now there are two control qubits. The target qubit will only be affected if both control qubits are $\ket{1}$.

\begin{figure}[h!]
\centering
\includegraphics{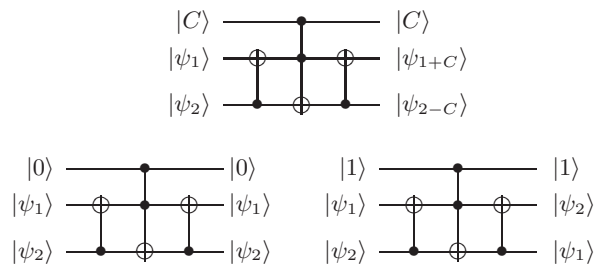}
\label{cswap}
\caption{Controlled qubit swap gate.}
\end{figure}

Superpositions are preserved in all these gates. Notice that the three gates are their own inverse. If we apply to a qubit any of them two times in a row we will recover the original state. Keeping that in mind we can see an important group of gates that acts as a conditional qubit swap. The quantum circuit of figure \ref{cswap} will swap the contents of two channels when the commutation qubit $\ket{C}$ is $\ket{1}$.

\section{Delayed commutation: Concept and limitations.}
\label{concept}
The inspiration for delayed commutation is the concept of a switch that sends to all its output ports a superposition of the states of each of the input ports. Ideally, when the routing decision has been taken, there is a mechanism that can reduce the state so that each of the destinations receives the information intended for it. 

We are assuming a network with two sources and two destinations. The information is sent in qubits in the general form $\alpha\ket{0}+\beta\ket{1}$. We suppose that the qubits from both of the sources arrive at the switch at the same time and there are no problems of contention, i.e. the sources will never ask for the same destination (we can assume that this is dealt with at a higher level).

In order to affect the content of the channel after the information has left the switch, we need to entangle the qubits that are travelling to the different destinations to some local ancillary qubits. Then, when the routing processing has been done, we can act on those ancillary qubits and reduce the state. The direct implementation of this idea is impossible. If we had two data sources, one emitting only $\ket{0}$ states and the other emitting only $\ket{1}$ states we could use the switch for faster than light information transmission. By deciding to switch or not the channels, we can send a $\ket{0}$ or $\ket{1}$ from the switch to the destination. This switch to destination communication could be faster than the speed of light. 

This gives us the first restriction. Part of the switching information must travel with the qubits. It seems possible to use teleportationlike schemes where the destination will apply different transformations on the received qubits depending on the values of a classical bit. In our proposal we prefer to use a modified form of superdense coding. The main advantage is that the routing information is part of the information qubits and we don't need additional qubits, or transmissions that will decrease the overall bit rate of the system. 

\section{Switch architecture for a simple network.}
\label{simple}

\begin{figure}[h!]
\centering
\includegraphics{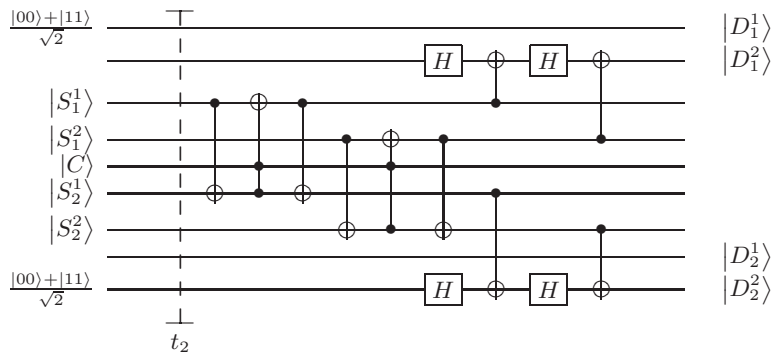}
\caption{Switch for delayed commutation.}
\label{switch}
\end{figure}

In this example, we take a network with two data sources $S_1$ and $S_2$, that can send packets to two destinations $D_1$ and $D_2$. We consider that the data from $S_i$ go to $D_i$ unless there is a ``commutation'' and we change the paths. We will study the simplest case, where the processing time is such that the routing information is not ready when the first qubit of the packet arrives, but it is known when the second one comes. The commutation will be controlled by the qubit $\ket{C}$. When  $\ket{C}=\ket{0}$, there is no commutation. When $\ket{C}=\ket{1}$ there is a commutation. With the circuit of figure \ref{switch} we can have delayed commutation.\\

We can see four ancillary qubits, two associated to each information source. Each pair starts in the entangled state $\frac{\ket{00}+\ket{11}}{\sqrt{2}}$. $|S_i^j\rangle$ is used to represent the qubit that the source $i$ emits at time $j$ (the $j$th qubit from the $i$th source). Thus qubits with different superindex arrive at the circuit at different moments. The $|D_i^j\rangle$ are the destination states following the same notation. 

We begin with some of the local ancillary qubits that are sent to the destination. When the first qubits from the sources arrive, they are kept in the switch, and the first qubits of the superpositions of ancillae are sent to $D_1$ and $D_2$. When the second qubit from each source arrives (at $t_2$ in the figure) we can use the doubly controlled NOT gates to switch the qubits or not depending on the value of $\ket{C}$. The second qubit defines the information that is transmitted. The information of the first qubit is carried in the phase ($0$ if it is $\ket{0}$, $\pi$ if it is $\ket{1}$). The information of the second qubit is going in the state of the transmitted qubit ($\ket{0}$ or $\ket{1}$). The logic sequence $\ket{ab}$, after the circuit, becomes $\frac{\ket{0b}+(-1)^a\ket{1b\oplus 1}}{\sqrt{2}}$.

Notice that there are two distinct parts in the circuit. The first one routes the qubits to the appropriate output port inside the switch. The second part (the two Hadamard gates, and the two CNOT gates following each one) takes the qubit into the superdense code acting on the phase and the state of the second qubit of the entangled ancillary state. 

\begin{figure}[h!]
\centering
\includegraphics{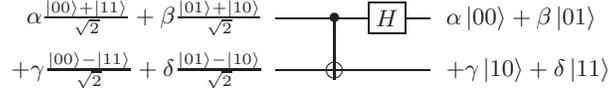}
\label{decoder}
\caption{Qbit decoder at the destination.}
\end{figure}

For the encoding 00: $\frac{\ket{00}+\ket{11}}{\sqrt{2}}$, 10: $\frac{\ket{00}-\ket{11}}{\sqrt{2}}$, 01: $\frac{\ket{01}+\ket{10}}{\sqrt{2}}$ and 11: $\frac{\ket{01}-\ket{10}}{\sqrt{2}}$, the received state will depend on whether we commute or not. If we use the system fot the transmission of classical information, i.e the sources only emit $\ket{0}$ or $\ket{1}$, we can measure the qubits that remain in the switch. If we use the network for quantum information transmission we need to store the original qubits in a quantum register until a measurement is made at the destination. The state will keep superpositions, but as the sent qubits are entangled to the qubits in the switch, changes in the local qubits would affect the transmitted information. We can recover the original encoding at the destination using a CNOT and a Hadamard gate (see Fig. \ref{decoder}).

\section{Generalization to complex networks.}
\label{general}
The result from the previous section can be extended to any number of inputs, outputs and number of qubits that arrive to the switch during the processing delay. The commutation we need to apply (following the control qubits) is defined as in \cite{TK02}. We just need to use a valid commutation circuit applied on the qubits going from the $(n/2+1)$-th to the $n$-th in the appropriate input lines. The superdense code we need to use will depend on the time it takes to get the routing information. In the usual form of superdense coding we can send two bits of classical information with just one qubit. In our case, that means that we can send $n/2$ qubits before knowing the destination. This is just a superdense coding scheme, with the advantage that it is not necessary to send the entangled qubits to the destination in advance. 

For instance, for a commutation delay of two qubits (the routing information is not available during the first two qubits), we will need a group of $2\times 2=4$ qubits. A valid code is:

\begin{table}[ht!]
\begin{center}
\begin{tabular}{cccccccc}	
0000:&$\frac{\ket{0000}+\ket{0101}+\ket{1010}+\ket{1111}}{2}$&0100:&$\frac{\ket{0000}-\ket{0101}+\ket{1010}-\ket{1111}}{2}$\\
1000:&$\frac{\ket{0000}+\ket{0101}-\ket{1010}-\ket{1111}}{2}$&1100:&$\frac{\ket{0000}-\ket{0101}-\ket{1010}+\ket{1111}}{2}$\\
0001:&$\frac{\ket{0001}+\ket{0100}+\ket{1011}+\ket{1110}}{2}$&0101:&$\frac{\ket{0001}-\ket{0100}+\ket{1011}-\ket{1110}}{2}$\\
1001:&$\frac{\ket{0001}+\ket{0100}-\ket{1011}-\ket{1110}}{2}$&1101:&$\frac{\ket{0001}-\ket{0100}-\ket{1011}+\ket{1110}}{2}$\\
0010:&$\frac{\ket{0010}+\ket{0111}+\ket{1000}+\ket{1101}}{2}$&0110:&$\frac{\ket{0010}-\ket{0111}+\ket{1000}-\ket{1101}}{2}$\\
1010:&$\frac{\ket{0010}+\ket{0111}-\ket{1000}-\ket{1101}}{2}$&1110:&$\frac{\ket{0010}-\ket{0111}-\ket{1000}+\ket{1101}}{2}$\\
0011:&$\frac{\ket{0011}+\ket{0110}+\ket{1001}+\ket{1100}}{2}$&0111:&$\frac{\ket{0011}-\ket{0110}+\ket{1001}-\ket{1100}}{2}$\\
1011:&$\frac{\ket{0011}+\ket{0110}-\ket{1001}-\ket{1100}}{2}$&1111:&$\frac{\ket{0011}-\ket{0110}-\ket{1001}+\ket{1100}}{2}$
\end{tabular}
\caption{Encoding for 4 qubits (processing delay of 2 qubits).}
\label{codificacion}
\end{center}
\end{table}
or, equivalently, $\ket{abcd}=\frac{\ket{00cd}+(-1)^b\ket{01cd\oplus 1}+(-1)^a\ket{10c\oplus 1d}+(-1)^{a+b}\ket{11c\oplus 1d\oplus 1}}{2}$.

In the general case of delay $n/2$ qubits we need a $n$-qubits code. We begin with the ancillary qubits in the state 
\begin{equation}
\frac{1}{\sqrt{2^{\frac{n}{2}}}}\sum_{i=0}^{2^{\frac{n}{2}}-1}\ket{i}\ket{i}.
\end{equation}
 Beginning from the $n/2$th qubit we can modify the qubits that leave the switch with a circuit like the one in Fig. \ref{switch}. The resulting code is given by the equation:
\begin{equation}
\ket{i}\ket{j}=\frac{1}{\sqrt{2^{\frac{n}{2}}}}\sum_{k=0}^{2^{\frac{n}{2}}-1}(-1)^{\vec{i}\cdot \vec{k}^\dag}\ket{k}\ket{j\oplus k},
\end{equation}
where $\ket{i}$ and $\ket{j}$ are the first and second half of the state vector, expressed as the binary representation of the decimal numbers $i$ and $j$. $j\oplus k$ is the bitwise XOR of those binary representations. Vectors $\vec{i}$ and $\vec{k}$ contain in each position the corresponding binary digit. It is easy to prove from this equation that we can generate an orthonormal basis so that every group of qubits in the traditional binary encoding is mapped into a new state.

\section{Analysis, discussion and future lines.}
\label{discuss}
Delayed commutation is a new application of quantum information that expands the capabilities of classical networks. With it we can avoid the processing delay at the nodes of the network. We have shown the quantum gates that those switches need.

The scheme uses a form of superdense coding. The code is given for $n$ qubits, and can compensate for a processing delay up to $n/2$ qubit time. As opposed to regular superdense coding schemes, we don't need to send the entangled qubits in advance. We only need to have in the switch $n$ ancillary qubits per channel. 

The improvement in the bit rate will depend on the transmission delay. The end-to-end delay without delayed commutation is $Dt_q+Nt_p$, where $D$ is the total distance from source to destination, $t_q$ the time it takes one qubit to cover one unit of space, $N$ the number of intermediate nodes, and $t_p$ the processing time at each node. We suppose that the operation time of the quantum gates in the switch is negligile when compared to the transmission and processing times. The bit rate will be $\frac{1}{Dt_q+Nt_p}$. In delayed commutation we eliminate the last term, so the bit rate is $\frac{1}{Dt_q}$. The improvement in the bit rate will be, $\frac{Dt_q+Nt_p}{Dt_q}=1+\frac{Nt_p}{Dt_q}$. This improvement is more important for networks with many intermediate nodes, a high processing time, and fast transmission times.

This model has one disadvantage. If we have more than one node the encoding will go from the superdense code to the regular code, and the other way round, in each switch. We can compensate that, either by granting an odd or even number of hops for each route (so that the destination knows in which encoding the data are), or by using a parity qubit that tells the number of hops. This will reduce the improvement in the bit rate to $\frac{Q_p}{Q_p+1}(1+\frac{Nt_p}{Dt_q})$, where $Q_p$ is the number of qubits of one packet. For a high number of bits in the packets the effects of this parity bit are negligible. We can also include a copy of the decoder circuit at the input of each of the switches. 

It is an open question what is the optimal code for delayed commutation. All the codes will give the same time improvement as the limit of the speed of light will remain. We call optimal code that with less ancillary qubits (needing less resources). As there are superdense coding variants that offer different information capacities \cite{SIM99,SIM99b} we can try to find the fittest one. Here it is also important to keep in mind the scalability of the solution.

Quantum information can be applied to new fields such as networking, with fruitful results. Delayed commutation is an example of a solution to a classical problem by means of quantum techniques. With it, the bit rate of future quantum networks can be substantially improved.

\bibliographystyle{apsrev}
\bibliography{lspace} 

\end{document}